\documentclass[a4paper]{spie}  
\usepackage[]{graphicx}

\title{A dedicated tool for a full 3D C$_N^2$ investigation} 

\author{Franck Lascaux\supit{a}, Elena Masciadri\supit{a} and Susana Hagelin\supit{a,b}
\skiplinehalf
\supit{a}INAF, Osservatorio Astrofisico di Arcetri, L.go E. Fermi 5, Florence, Italy; \\
\supit{b}Uppsala University, Department of Earth Sciences, Villav\"agen 16, 752 36 Uppsala, Sweden
}

\authorinfo{Further author information: (Send correspondence to Elena Masciadri)\\Elena Masciadri: E-mail: masciadri@arcetri.astro.it, Telephone: +39 055 2752 203}

\begin{document} 
  \maketitle 

\begin{abstract}
We present in this study a mapping of the optical turbulence (OT) above different astronomical sites.
The mesoscale model Meso-NH was used together with the Astro-Meso-Nh package and a set of diagnostic tools 
allowing for a full 3D investigation of the C$_N^2$.
The diagnostics implemented in the Astro-Meso-Nh, allowing for a full 3D investigation of the OT structure in a 
volumetric space above different sites, are presented.  
To illustrate the different diagnostics and their potentialities, we investigated one night and looked at instantaneous fields of 
meteorologic and astroclimatic parameters.
To show the potentialities of this tool for applications in an Observatory we ran the model above sites with very 
different OT distributions: the antarctic plateau (Dome C, Dome A, South Pole) and a mid-latitude site (Mt. Graham, Arizona).
We put particular emphasis on the 2D maps of integrated astroclimatic parameters (seeing, isoplanatic angles) 
calculated in different slices at different heights in the troposhere. 
This is an useful tool of prediction and investigation of the turbulence structure.
It can support the optimization of the AO, GLAO and MCAO systems running at the focus of the ground-based telescopes.
From this studies it emerges that the astronomical sites clearly present different OT behaviors.
Besides, our tool allowed us for discriminating these sites. 
\end{abstract}


\keywords{Optical Turbulence, Site Testing, Numerical Simulations}

\section{INTRODUCTION}
\label{sec:intro}  
Our team has already proven that the mesoscale model Meso-NH \cite{Lafore98} performed better than the European Centre 
for Medium-range Weather Forecasts (ECMWMF) General Circulation Model (GCM) in reproducing the meterological parameters 
which the optical turbulence depends on (wind speed, temperature), above different parts of the Antarctic Plateau \cite{Hagelin08}. 
A study by our team \cite{Hagelin10} also proved that the Meso-NH model can provide good nightly estimates of the wind distribution on 
the vertical at the Mount Graham astronomical site from the ground up to the top of the atmopshere (around 20 km).
The Astro-Meso-NH package \cite{Masciadri99a} implemented in the Meso-NH model was proved to be able to reconstruct realistic C$_N^2$ 
profiles above astronomical sites \cite{Masciadri99b,Masciadri01a} and was statistically validated above mid-latitude 
sites \cite{Masciadri01b, Masciadri04, Masciadri06} and more recently at Dome C, Antarctica \cite{Lascaux09, Lascaux10}.
\\
In this study, we intend to demonstrate the potentiality of such a tool in describing the OT above different astronomical 
sites.
We do not deduce from this study any statistical conclusion about the OT. The goal is to illustrate how our mesoscale model and  
some features of its associated diagnostic tools can be used to fully investigate a volumetric space above a given site, 
having access to the 3D C$_N^2$ and the associated astroclimatic parameters.
We focused our study on three antarctic sites (Dome C, Dome A and South Pole) and one mid-latitude site (Mount Graham, Arizona). 
All have clearly observed different OT behaviours.
\\
We put particular emphasis on the 2D maps of integrated astroclimatic parameters like seeing and isoplanatic angles, calculated 
for different slices of the troposphere.
This is very useful for the prediction and the investigation of the turbulence structure.
More over, it can support the optimization of the AO, GLAO and MCAO systems running at the focus of the ground-based telescopes.
\\
The first section is dedicated to a brief overview of the numerical setup.
The second section presents vertical cross-sections of C$_N^2$ from the ground up to around 800 m above the ground, and from the 
ground up to 20 km above see level, for a given night.
In the third section, we investigate some integrated astroclimatic parameters (seeing and isoplanatic angle).
\section{Numerical configuration}
\begin{figure}
   \begin{center}
   \begin{tabular}{c}
   \includegraphics[width=\textwidth]{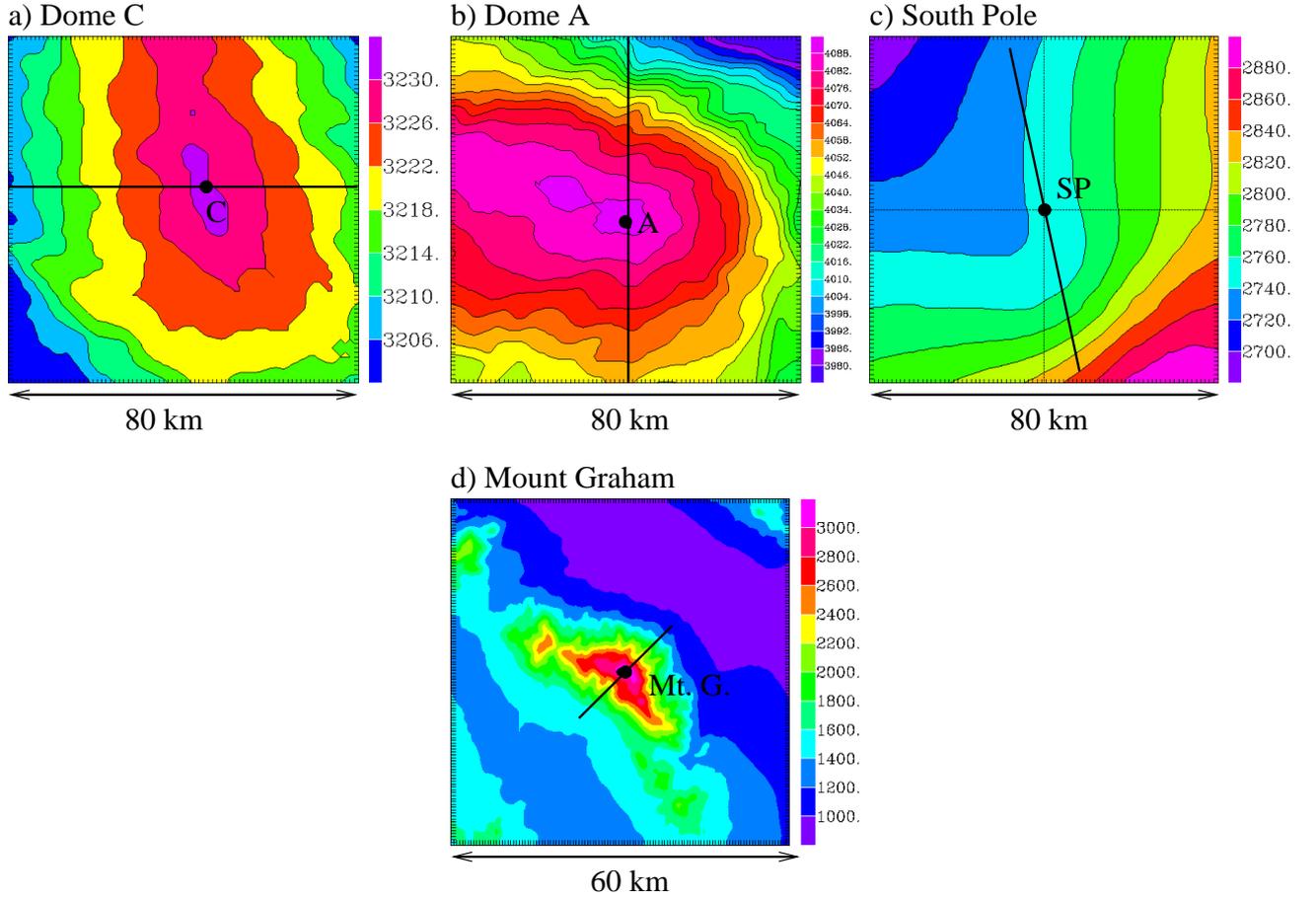}
   \end{tabular}
   \end{center}
   \caption[example]
   { \label{fig:zs}
Orography of a/ Dome C area (isolevels every 4 m, first at 3206 m), b/ Dome A area (isolevels every 6 m, first at 3980 m, 
c/ South Pole area (isolevels every 15 m, first at 2700 m) and d/ Mt. Graham area (isolevels every 200 m, first at 1000 m). Units in m. 
Black lines represent the positions of the vertical cross sections of the next figures.}
   \end{figure}

In this study we used the Meso-NH mesoscale model \cite{Lafore98} together with the Astro-Meso-NH package \cite{Masciadri99a}.
For each investigated antarctic site (Dome A, Dome C and South Pole) we used the same configuration to forecast the same winter night.
A grid-nested configuration \cite{Stein00} with three imbricated domains was employed. 
The three domains have increasing horizontal resolutions (mesh-sizes of 25 km, 5 km and 1 km, respectively) and the same vertical grid.
The vertical grid is a stretched Gal-Chen \& Sommerville \cite{Gal75} grid with the first point at 2 m above the ground and 12 points 
in the first 100 m. 
Above 3.5 km the vertical resolution is constant and equal to about 600 m.
More about the configuration can be found in recent publications from our group \cite{Lascaux09,Lascaux10}.
The configuration of the simulation at Mount Graham is slightly different.
The grid-nesting configuration is still made of three domains, but the horizontal resolutions are different: 10 km, 2.5 km and 0.5 km, 
respectively.
The vertical grid is also different: it is also a stretched grid, but the first point is at 20 m.
All simulations are initialized and forced with the analyses from the ECMWF (European Centre for Medium-Range Weather Forecast).
In this study we will consider only the fields computed inside the innermost domain for each of the four sites.
On figure \ref{fig:zs} is shown the orography for all the studied sites. 
The black lines seen on every figure are the lines along which are done the vertical cross-sections of C$_N^2$ analysed in the next 
sections.

\section{Vertical cross-sections of C$_N^2$} 
\begin{figure}
   \begin{center}
   \begin{tabular}{c}
   \includegraphics[width=\textwidth]{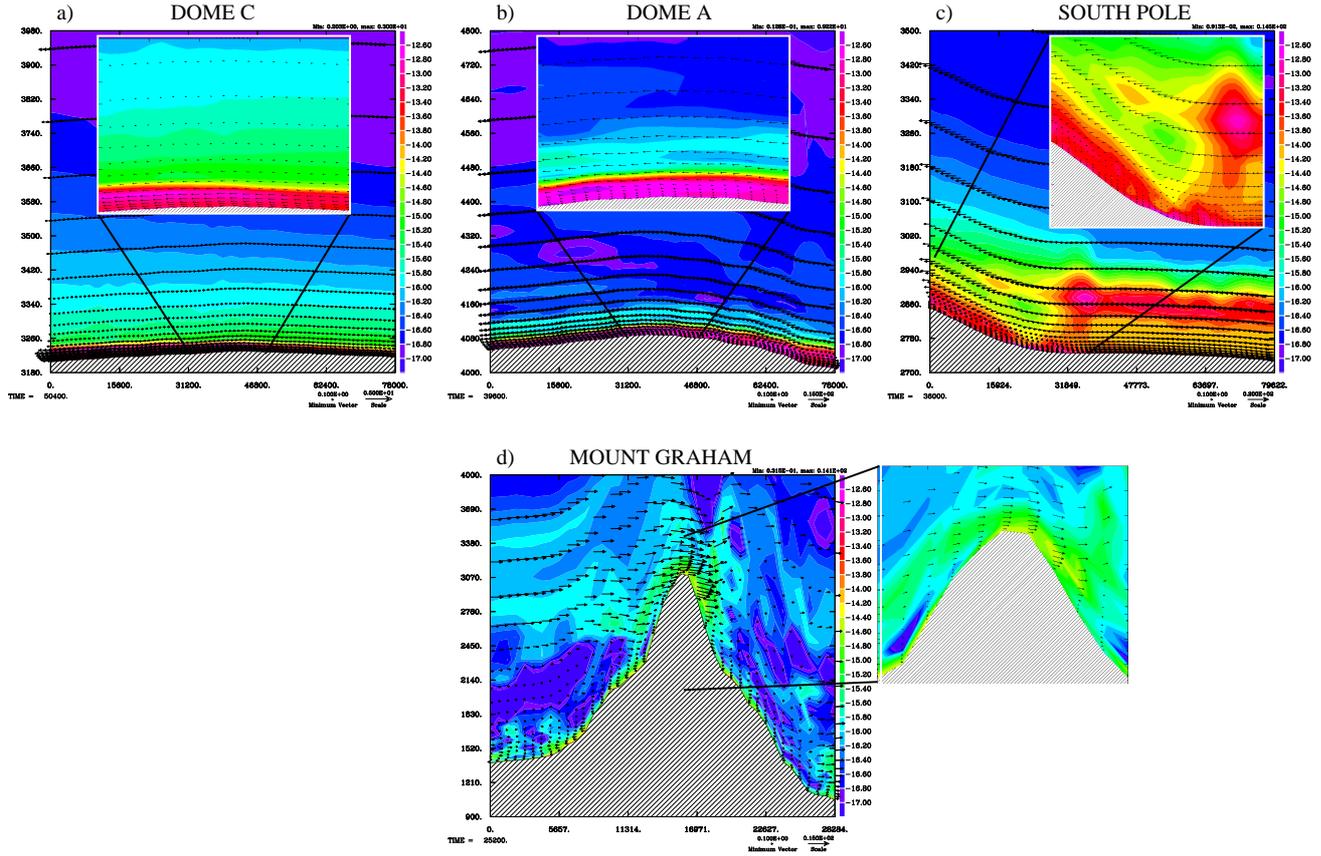}
   \end{tabular}
   \end{center}
   \caption[example]
   { \label{fig:cv_800m}
Vertical cross section of C$_N^2$ (first 800 m of the atmosphere) along the horizontal lines displayed in Figure \ref{fig:zs} 
a/ at Dome C, b/ at Dome A, c/ at South Pole (at 22 LT - same night for the three antarctic sites) and d/ at Mount Graham (at 00 LT). 
Logarithimic unit in m$^{-2/3}$. Superimposed is the wind inside the cross section (in m.s$^{-1}$.}
   \end{figure}

\begin{figure}
   \begin{center}
   \begin{tabular}{c}
   \includegraphics[width=\textwidth]{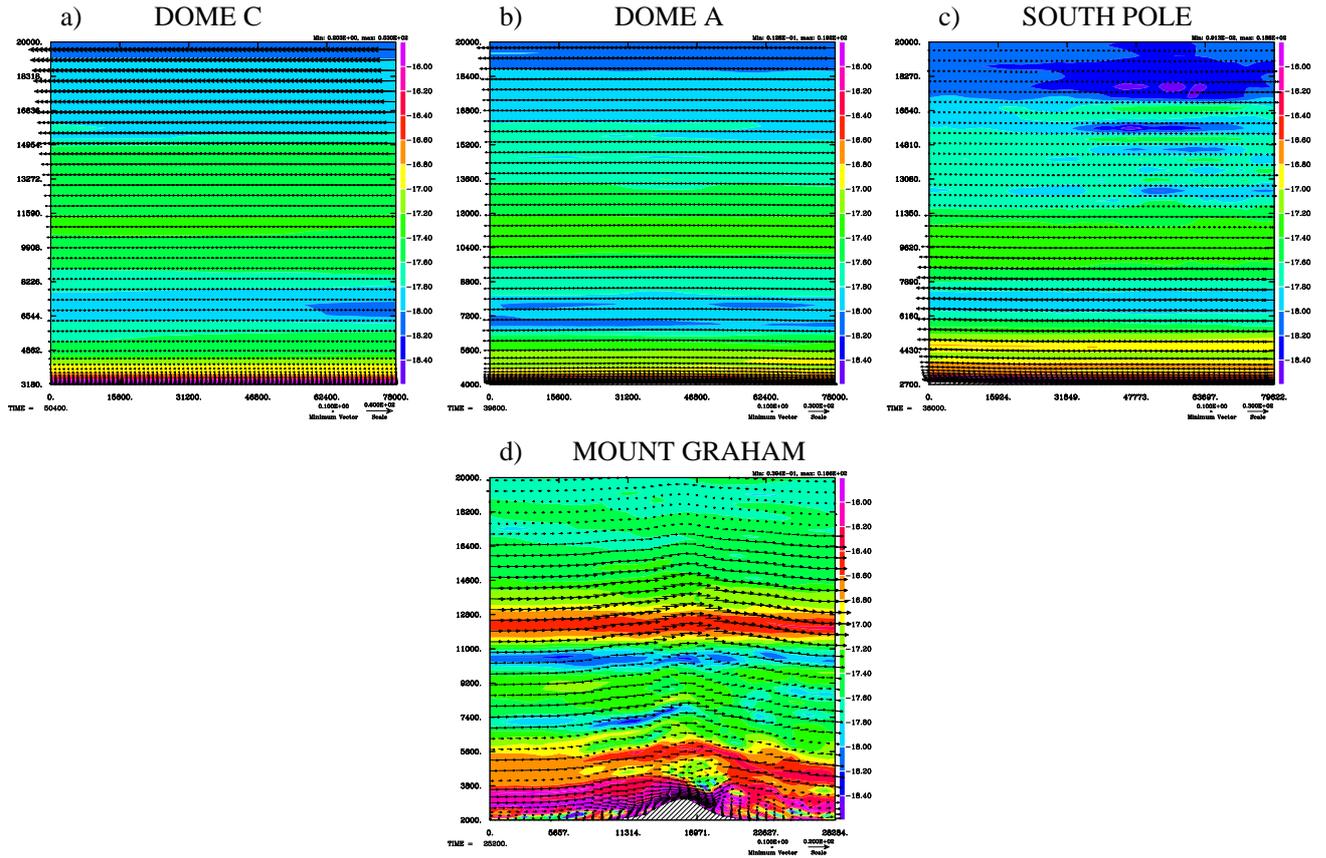}
   \end{tabular}
   \end{center}
   \caption[example]
   { \label{fig:cv_20km}
Same as Figure \ref{fig:cv_800m} but up to 20 km (above see level).}
   \end{figure}

In this section we present one of the possibilities of investigation of the OT fields provided by the diagnostic tools of the 
Astro-Meso-NH package.
In Figures \ref{fig:cv_800m} and \ref{fig:cv_20km} are shown an example of the vertical cross-sections of the C$_N^2$ along the 
lines represented in Figure \ref{fig:zs} from the ground up to around 800 m and from the ground up to 20 km, respectively, 
for all the investigated sites.
These are instantaneous cross-sections of one simulated night.
The winter night simulated for illustrating our tool is the same for the antarctic sites (Dome C, Dome A and South Pole) and 
the plots are made at 22 LT.
The night simulated at Mount Graham is a different night, but still in winter. 
Instantaneous plots are extracted at 00 UT.
We remind the reader that these plots are illustrative and we do not claim to infer any statistical behaviours from them.
\\
In Figure \ref{fig:cv_800m} it is clearly put into evidence different patterns in the vertical distribution of the OT in the 
first part of the atmosphere.
Either Dome C and Dome A have the OT concentrated in the first tenths of meters as evidenced in Figures \ref{fig:cv_800m}a,b.
For this night, we can see that the OT is even more concentrated at Dome A (where Log(C$_N^2$) reached the value of -16 m$^{-2/3}$ 
at 100 m above the ground) than at Dome C (where Log(C$_N^2$) reached the value of -16 m$^{-2/3}$ at only 250 m).
The typology of the vertical distribution of the OT at South Pole is a bit different for this night.
Indeed, we can clearly see a first maximum of OT near the ground (just like for Dome A and Dome C) and a second maximum above, near around 
150 m above the ground.
Also interesting to notice is the establishment of a katabiatic wind along the slopes in the first meters.
At Mount Graham, a mid-latitude site, the distribution of the OT is different as it can be seen in Figure \ref{fig:cv_800m}d.
The wind field is disturbed by the mountain, and a wave is forming right after the peak.
At the peak the maximum of the OT, near the ground, is around -14.5 m$^{-2/3}$ (Log(C$_N^2$)), whereas at Dome A (where it was 
more concentrated near the ground) the maximum reached -12.5  m$^{-2/3}$, thus much higher.
The OT at Mount Graham was spread over a deeper layer than above the antarctic sites.
\\
We can also investigate the vertical distribution of the C$_N^2$ for all the troposphere.
This is displayed in Figure \ref{fig:cv_20km}, same as Figure \ref{fig:cv_800m} but until 20 km this time.
For the night considered, above 1 km above the ground level, the vertical distributions at Dome C, Dome A and South Pole are 
quite undisturbed, and very similar.
A minimum is reached at around 4 km above the ground (5 km for South Pole) with values of Log(C$_N^2$) close to -18 m$^{-2/3}$.
From 6 or 7 km up to the top it regularly decreases in the same manner for the three sites.
The wind speed in altitude is much higher at Dome C (maximum of 53 m.s$^{-1}$) than at Dome A (maximum of 19 m.s$^{-1}$) and 
South Pole (less than a few m.s$^{-1}$ in altitude) for this night. 
This is certainly explained by the presence of the vortex in altitude \cite{Hagelin08,Lascaux10b}.
\\
The vertical distribution of the OT is different at Mount Graham.
Due to the presence of a jet stream centered at 12 km above sea level, at the difference of an antarctic site, 
a secondary maximum of OT is visible (with values for Log(C$_N^2$) close to -16  m$^{-2/3}$).
It is well evidenced in Figure \ref{fig:cv_800m}d.
Another difference is in the first kilometers above the ground at Mount Graham: they are characterized by an OT spread over a 
deeper vertical layer than in Antarctica.
\\
Now let's have a look at some integrated astroclimatic parameters.
\section{Integrated astroclimatic parameters} 
All the integrated astroclimatic parameters that are of interest for astronomers (seeing, isoplanatic angle, wavefront coherence time, 
spatial coherence outer scale) are accessible with our model.
Here we present examples of 2D maps of seeing and isoplanatic angle.
\subsection{Horizontal maps of seeing} 
\begin{figure}
   \begin{center}
   \begin{tabular}{c}
   \includegraphics[width=0.8\textwidth]{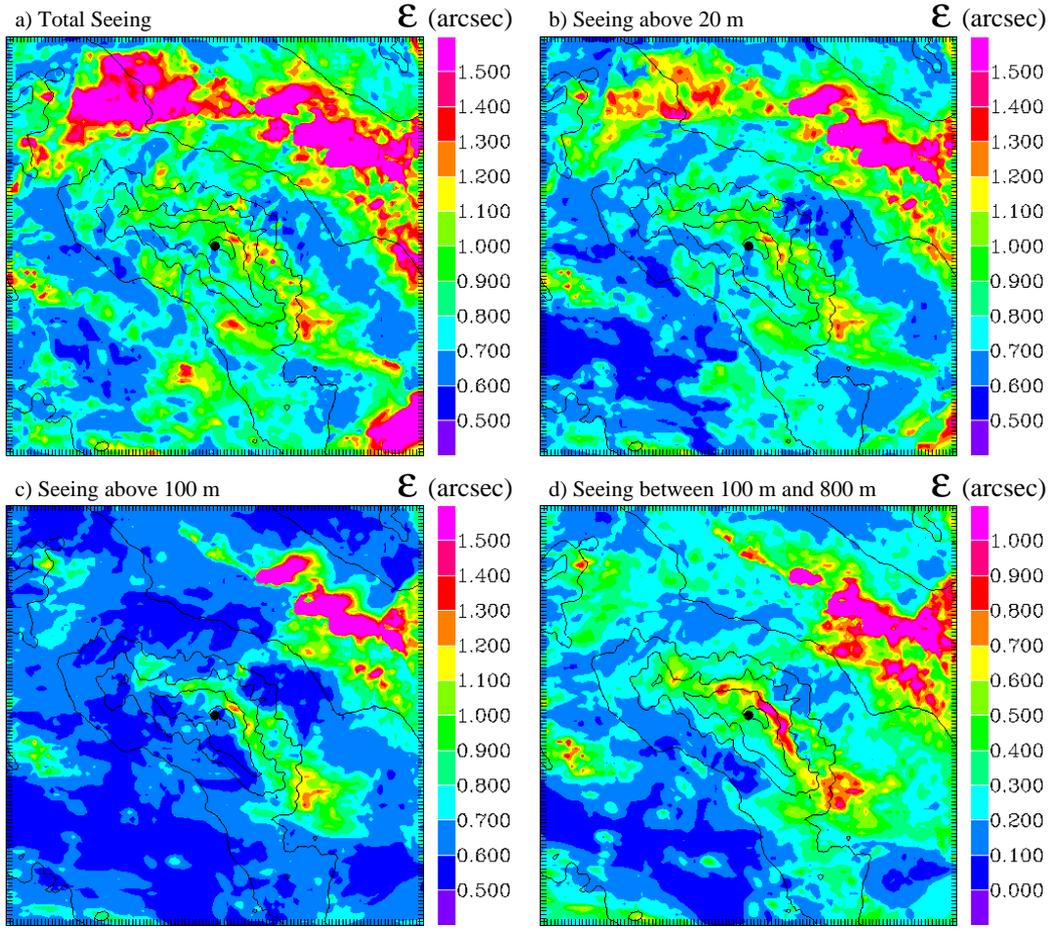}
   \end{tabular}
   \end{center}
   \caption[example]
   { \label{fig:ch_seeing}
Horizontal computed maps of seeing at Mount Graham. a/ Total seeing (from the ground up to the top of the model), b/ Seeing above 20 m 
(up to the top of the model), c/ Seeing above 100 m (up to the top of the model) and d/ Seeing between 10 km and 13 km above the ground.
Unit in arcsec. The black dot is the location of the Mount Graham osservatory.}
   \end{figure}

Figure \ref{fig:ch_seeing} shows two-dimensional maps of seeing at Mount Graham, for the same night than in the previous section.
Not only we can access the total seeing (i.e from the ground to the top of the model), but also the contibution to the seeing of any 
given slice of the atmosphere.
Here we chose to display the total seeing (Fig. \ref{fig:ch_seeing}a), the seeing above 20 m (Fig. \ref{fig:ch_seeing}b), 
the seeing above 100 m (Fig. \ref{fig:ch_seeing}c) and the contribution to the seeing of the atmosphere between 100 m and 800 m. 
All the altitude previously mentioned are above ground level.
\subsection{Horizontal maps of isoplanatic angle}
\begin{figure}
   \begin{center}
   \begin{tabular}{c}
   \includegraphics[width=\textwidth]{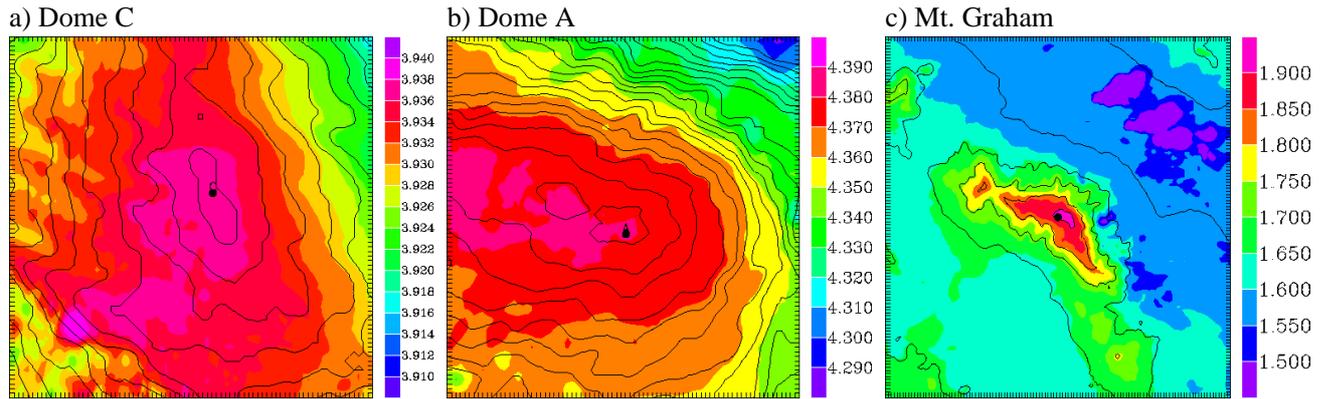}
   \end{tabular}
   \end{center}
   \caption[example]
   { \label{fig:ch_iso}
Horizontal computed maps of isoplanatic angle a/ at Dome A, b/ Dome C and c/ Mt. Graham. Unit in arcsec.}
   \end{figure}

Another integrated parameter derived from the C$_N^2$ is presented.
On Figure \ref{fig:ch_iso} one can see the 2D maps of the instantaneous isoplanatic angle for three different sites: Dome C, Dome A and 
Mount Graham.
The values of the isoplanatic angle are quite uniform over the entire domain for Dome A and Dome C, with higher values for Dome A 
($\Theta{_{DA}}$=4.38 arcsec and $\Theta{_{DC}}$=3.94 arcsec).
At Mount Graham, the isoplanatic angle is obviously much more dependent on the orography. 
The highest value is reached at the peak: $\Theta{_{MtG}}$=1.9 arcsec, and is half the value at Dome C or Dome A.

\section{Conclusion}
In this study we illustrated how the mesoscale model Meso-NH and the associated Astro-Meso-NH package can be use to investigate the OT 
at a given site.
One of the major advantage of such a tool is the possibility to study a full three-dimensional C$_N^2$, and the associated 
integrated astroclimatic parameters.
Here we presented some of the features useful for a complete mapping of the OT:
\begin{itemize}
\item vertical cross-sections of C$_N^2$;
\item 2D maps of total seeing;
\item 2D maps of the contribution to the seeing of any desired slice of the atmosphere;
\item 2D maps of isoplanatic angles.
\end{itemize}
Others parameters can be studied, even if they were not shown here, like the wavefront coherence time or the spatial coherence outer scale.
Using this tool, we can easily deeply investigate the OT at given sites and discriminate between them.
In our examples, it is clearly put into evidence that for the chosen night, the OT was more concentrated in the first tenth of meters 
above the ground at Dome C and Dome A, whereas a second layer around 150 m was visible at South Pole.
At Mount Graham, for the night considered in this study, the intensity of the optical turbulence near the ground is less 
strong than in the antarctic sites, but it is distributed over a deepest layer of the atmosphere.
\\
Obviously others features are available, but were not presented in this study.
The reader can refer to the website of our group {\it http://forot.arcetri.astro.it./} for more examples.


\acknowledgments     
The study has been funded by the Marie Curie Excellence Grant (FOROT)-MEXT-CT-2005-023878.
ECMWF products are extracted by the MARS catalog
http://www.ecmwf.int and authors are authorized to use
them by the Meteorologic Service of the Italian Air Force.


\end{document}